\documentclass[twocolumn,pre,aps,floatfix,amsmath,amssymb,superscriptaddress,longbibliography]{revtex4-2}

\usepackage{xcolor}
\usepackage{hyperref}

\usepackage{graphicx}
\usepackage{subfigure} 
\usepackage{physics}

\usepackage{amsthm}
\usepackage{amsmath}
\usepackage{amssymb}
\usepackage{calligra}
\usepackage{dsfont} 

\usepackage{calligra}
\DeclareMathAlphabet{\mathcalligra}{T1}{calligra}{m}{n}
\DeclareFontShape{T1}{calligra}{m}{n}{<->s*[2.2]callig15}{}

\DeclareSymbolFont{usualmathcal}{OMS}{cmsy}{m}{n}
\DeclareSymbolFontAlphabet{\mathcal}{usualmathcal}

\usepackage{tikz} 
\usetikzlibrary{arrows,scopes} 

\usepackage[color=orange!60,textsize=scriptsize]{todonotes}
\usepackage{siunitx}
\DeclareSIUnit\angstrom{\text{}} 

\usepackage{pdfpages} 
\usepackage{pgffor} 
\usepackage{xr} 

\makeatletter
\AtBeginDocument{\let\LS@rot\@undefined}
\makeatother

\newcommand{\N}{\mathbb{N}}

\newcommand{\R}{\mathbb{R}}
\newcommand{\C}{\mathbb{C}}

\newcommand{\RR}{\mathcal{R}}

\newcommand{\QQ}{\mathcal{Q}}

\renewcommand{\AA}{\mathcal{A}}
\renewcommand{\SS}{\mathcal{S}} 

\newcommand{\diff}{\mathrm{d}} 
\newcommand{\Diff}{\mathcal{D}}

\newcommand{\ii}{\mathrm{i}}
\newcommand{\ee}{\mathrm{e}}

\newcommand{\oneq}{\mathds{1}_{q}}
\newcommand{\onep}{\mathds{1}_{p}}

\newcommand{\inv}[1]{#1^{-1}}
\newcommand{\tran}[1]{#1^{\top}}

\newcommand{\comma}{\ , \qquad}
\newcommand{\qiff}{\quad \iff \quad}

\newcommand{\avg}[1]{\langle #1 \rangle}
\newcommand{\Avg}[1]{\left\langle #1 \right\rangle}

\newcommand{\Abs}[1]{\left\vert #1 \right\vert}

\makeatletter
\newcommand{\vast}{\bBigg@{3}}
\newcommand{\Vast}{\bBigg@{4}}
\newcommand{\VAST}{\bBigg@{5}}
\makeatother

\renewcommand{\Im}[0]{\operatorname{Im}}

\newcommand{\adm}[1]{{#1}}

\begin{document}

\title{BBP Phase Transition for an Extensive Number of Outliers}

\author{Niklas Forner}
\affiliation{Institut f\"ur Theoretische Physik, Universit\"at Leipzig, D-04103, Leipzig, Germany}

\author{Alexander Maloney}
\affiliation{Department of Physics, Syracuse University, New York, USA}
\affiliation{Institute for Quantum \& Information Sciences, Syracuse University, New York, USA}
\affiliation{Department of Physics, McGill University, Montr{\'e}al, Canada}

\author{Bernd Rosenow}
\affiliation{Institut f\"ur Theoretische Physik, Universit\"at Leipzig, D-04103, Leipzig, Germany}

\begin{abstract} 

Random-matrix theory helps disentangle signal from noise in large data sets. 
We analyze rectangular $p\times q$ matrices $W = W_0 + M$ in which the noise $M$ generates a Marchenko–Pastur bulk, whereas the signal $W_0$ injects an extensive set of degenerate singular values. 
Keeping $\rank (W_0) /q$ finite as $p,q \to \infty$, we show that the trace of the resolvent of $W^{\top} W$ obeys a quartic equation for one degenerate signal, yielding an exact spectral density, and derive explicit asymptotics in the strong-signal regime.
We map out a detailed generalized Baik–Ben Arous–Péché (BBP) phase diagram and clarify how a finite density of spikes reshapes the bulk edges.
We further derive a $1/3$-scaling law for the critical signal strength in terms of the rank ratio for rectangular matrices in the finite-to-extensive-rank crossover.
Numerical simulations validate the theory and illustrate its relevance for high-dimensional inference tasks with multiple degenerate signals and more general signal distributions. 


\end{abstract}

\maketitle

\section{Introduction} 

Contemporary data analysis frequently relies on large data sets -- for example, multivariate time-series measurements or the weight matrices of deep neural networks. 
These data are invariably contaminated by noise. 
Although the main goal is to uncover the underlying signal, it is equally important to understand how noise acts on that signal to produce the observations \cite{Bendat2010}.
Noise can take many forms and may exhibit correlations between different data points, thereby modifying the relation between the latent signal and the measured data.

Random Matrix Theory (RMT) was developed to predict statistical properties -- such as eigenvalue and singular-value distributions -- of large matrices whose entries are random variables. 
The framework has found applications spanning statistical and nuclear physics \cite{Wigner-55-Random,Wigner-58-Random}, econophysics and finance \cite{Stanley-00-Econophysics,Laloux-99-Finance}, quantum mechanics \cite{Guhr-98-QM, Weidenmueller-2024-Thermalization}, image processing \cite{Olesen-23-Medicine}, wireless communications \cite{Sengupta-00-Correlations}, neural networks \cite{Lampinen-19-Network, Mahoney-2021-Networks, Thamm-2022-RMT}, and many other fields.

The spectra of random matrices have been studied extensively, and numerous models for signal matrices perturbed by noise have been proposed, see Refs.~\cite{Bai-99-Random,Bun-17-Report} for reviews. 
In modern applications these matrices are often rectangular, and the dominant noise is typically additive -- for instance, fluctuations of weights around a training minimum.
In this work we consider rectangular $p\times q$ ``signal-plus-noise’’ matrices $W = W_0 + M$, where $W_0$ encodes the signal and $M$ represents the noise, here realized by independent and identically distributed (i.i.d.) entries. 
Whereas a square matrix is characterized by its eigenvalues, a rectangular matrix is described by the eigenvalues of $W^{\!\top} W$, i.e., by the squared singular values of $W$. 
Quantifying how noise reshapes this singular-value spectrum is essential for inferring the true signal weights in settings such as late-stage neural-network training \cite{Lampinen-19-Network}.
In the \adm{``thermodynamic''} limit $p,q\to\infty$ with $p/q$ fixed and finite the spectral density is self-averaging\adm{, in the sense that the statistical properties of a typical matrix approach those of the ensemble as a whole} \cite{Bun-17-Report,Benaych-12-Singular}. 
The noise component (the ``bulk’’) forms a continuous density with sharp edges \cite{Tao-13-Outliers,Bai-98-Sharp,BBP-05-Transition}. 
A low-rank perturbation satisfying $\mathrm{rank}\,W_0/q\to 0$ does not affect the bulk -- the Marchenko-Pastur law \cite{MP-67-Distributions} -- but produces isolated eigenvalues.
\adm{When these eigenvalues are large enough, they will separate from the continuous bulk; the point at which they emerge from the bulk is} the Baik–Ben Arous–Péché (BBP) phase transition \cite{Peche-05-Largest,BBP-05-Transition}.

\begin{figure*}[t]
    \subfigure[$\vartheta^2 = 0.2$]{
        \includegraphics[width=0.31\linewidth]{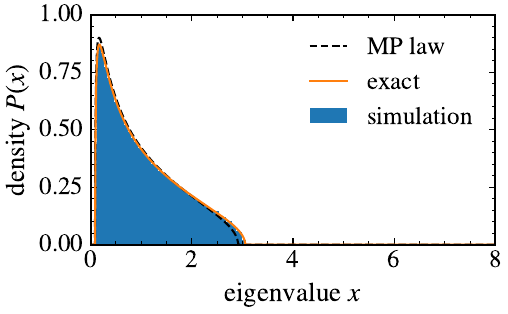}
        \label{subfig:spectrum-splitting-1}
    }
    \hfill
    \subfigure[$\vartheta^2 = 1.82$]{
        \includegraphics[width=0.31\linewidth]{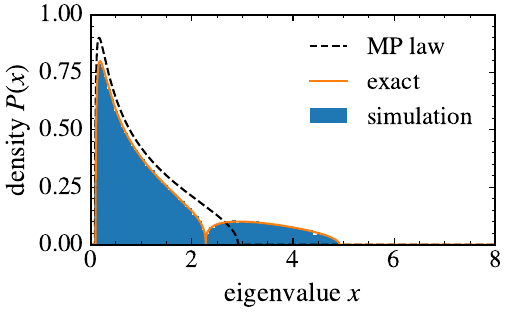}
        \label{subfig:spectrum-splitting-2}
    }
    \hfill
    \subfigure[$\vartheta^2 = 4$]{
        \includegraphics[width=0.31\linewidth]{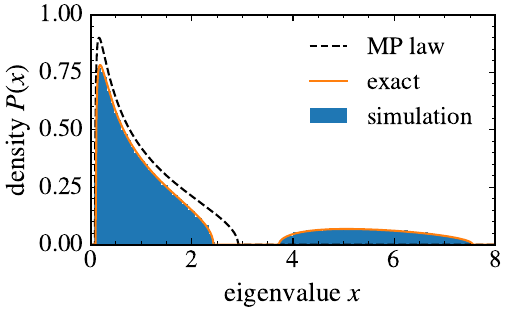}
        \label{subfig:spectrum-splitting-3}
    }
    \caption{Splitting of the spectrum for increasing signals at a fixed rank ratio. $\sigma = 1$, $\AA = 2$, $\tilde{r} = 0.2$. 
    The simulations for the spectrum of $\tran{W} W$ are conducted for $p \times q$ matrices $W = M + W_0$ with $p = \AA q$ and $q = 1000$, averaged over a number of ten runs. 
    Each entry in $M$ is drawn from a centered normal distribution with fixed variance $\sigma^2 / p$. The Marchenko-Pastur (MP) law acts as a reference for the noise-only case.}
    \label{fig:spectrum-splitting}
\end{figure*}

Here we analyze the complementary regime, an {\em extensive} number of outliers \adm{-- i.e. a signal matrix whose rank scales linearly with $p$ and $q$ --} whose collective effect reshapes the bulk spectrum (see Fig.~\ref{fig:spectrum-splitting}). 
\adm{We find a similar phase transition, where for sufficiently large signal an entire continuous signal distribution separates from the bulk ``noise'' part of the spectrum. 
Of course, the location of this phase transition, as well as the shape of both the bulk and the signal distributions, will depend on the properties of $W_0$ and $M$}. 
We begin with a single deviating singular value of degeneracy $r$, with rank ratio ${\tilde r} := r/q$ finite in the large $q$ limit. 
The full spectrum is then governed by a fourth-order algebraic equation, from which we \adm{determine the location of the phase transition, the shape of the signal and noise distributions, and} we derive an explicit asymptotic solution at large signal strength.
We map out the generalized BBP phase diagram as a function of ${\tilde r}$ and signal strength, and uncover a scaling law relating the critical signal strength to both ${\tilde r}$ and the aspect ratio $p/q$.
We validate these analytical results \adm{by matching to} numerical simulations. 
\adm{For more complicated signal distributions with $n$ distinct singular values, we find an analogous result where the spectrum is determined by an algebraic equation of order $2n+2$.}

A broad literature addresses so-called spiked covariance models, beginning with Johnstone \cite{Johnstone-2001-PCA} and followed by the characterization of the BBP phase transition \cite{BBP-05-Transition, Baik-2006-Spiked} using free probability theory \cite{Ryan-2007-Convolution, Benaych-12-Singular}.
Here, we formulate the problem in the signal-plus-noise setting using the replica approach.
The replica method, introduced by Edwards \cite{Edwards-1976-Replica}, found application in these types of setups \cite{Sengupta-99-Distributions, Hoyle-03-Outliers, Hoyle-04-PCA}.
The results below show that the methods yield the same self-consistency equations for the resolvent of $W^{\top} W$, or rather its trace \cite{Silverstein-95-Existence, Dozier-07-Green}, and that, at least in the i.i.d. case, the spiked-covariance and signal-plus-noise models are equivalent.
The comprehensive literature on finite-rank signals is complemented by recent work on the large-rank regime, for instance, on edge statistics in multiplicative models \cite{Ding-22-Statistics, Ding-2023-Multiplicative} or spectral dynamics during network training \cite{Lauditi-2026-Dynamics}.

In related work, Ganguli et al.~\cite{Ganguli-23-Extensive} analyze an extensive-rank ``spike’’ model in the context of optimal, rotationally invariant estimators for the population covariance matrix, with further developments in Refs.~\cite{Pourkamali-2023-Estimators,Troiani-2022-Denoising,Barbier-2022-Estimators,Maillard-22-Denoising,Fleig-22-Denoising, Pourkamali-2023-Priors, Pourkamali-2024-Inference, Barbier-2025-Denoising}. 
Our analysis addresses the same regime and independently reproduces their principal findings on the spectral splitting within a complementary formulation, while extending them in several significant ways.
First, we uncover a scaling law that relates the critical signal strength to both the rank ratio and the matrix aspect ratio in the crossover from finite-rank to extensive-rank signals.
Second, we derive an explicit asymptotic expression for the signal-induced part of the spectrum, which avoids the numerical solution of self-consistency equations and remains accurate for finite matrix sizes. 
Third, for signal ensembles with power-law distributed singular values, we obtain a closed analytic equation for the spectral function, allowing a detailed characterization of the corresponding population distribution.

\section{Model} 
We consider a real $p \times q$ matrix expressed as a sum of signal and noise $ W = W_0 + M$ with $p, q \in \N$. 
Our goal is to find the limiting singular–value density $\rho(\lambda)= 2\lambda\,P(\lambda^{2})$, where $P(x)$ is the average eigenvalue density of $W^{\!\top} W$ in the thermodynamic limit $p,q\to\infty$ at fixed aspect ratio $\mathcal A := p / q$.
Here, we let $\AA \geq 1$ to forego the $q-p$ vanishing singular values that would otherwise be present. 
Averages $\langle\cdot\rangle_{M}$ are taken with respect to the Gaussian measure
\begin{align}\label{eq:Gaussian-noise-integration} 
    \langle \, \cdot \, \rangle_{M} &= \frac{1}{(2 \pi \sigma^2 / p)^{pq/2}} \int_{\mathbb{R}^{pq}} \mathcal{D} M \ (\, \cdot \,) \, \mathrm{e}^{- \frac{p}{2 \sigma^2} \, \mathrm{Tr} (M^{\top} M)} \ 
\end{align}
where $\Diff M = \prod_{i = 1}^{p} \prod_{j = 1}^{q} \diff M_{ij}$. 
The  entries $M_{ij}$ are independent and identically distributed (i.i.d.) random variables with zero mean:
\begin{align}\label{eq:Gaussian-noise-moments} 
    \avg{M_{ia}}_{M} = 0 \comma \langle M_{ia} M_{jb} \rangle_{M} = \frac{\sigma^2}{p} \, \delta_{ij} \delta_{ab} \, .
\end{align}
We introduce the Green function as the limiting average trace of the resolvent of $\tran{W} W$ 
\begin{align}
    G(z) :=& \, \lim_{q \to + \infty} \bigg\langle \frac{1}{q} \Tr \left[ \frac{1}{z - \tran{W} W} \right] \bigg\rangle_{M} \label{eq:def-Green-trace}
\end{align}
which is normalized so that the eigenvalue density is  
\begin{equation}\label{eq:setup-spectrum}
    P(x) = - \frac{1}{\pi} \Im G(x  + \ii 0^{+}) \, .
\end{equation}
We further define the partition function $Z$ by the following multidimensional Gaussian integral,
\begin{subequations}
    \begin{align}
        Z(z) :=& \, \frac{1}{\sqrt{\det (z  - \tran{W} W)}} \label{subeq:def-partition-determinant} \\
        =& \, \biggl( \frac{q}{2 \pi} \biggr)^{q/2} \int_{\R^{q}} \Diff X \ \ee^{- \frac{q}{2} \tran{X} (z  - \tran{W} W) X} \label{subeq:def-partition-integral}
    \end{align} 
\end{subequations}
with $\Diff X = \prod_{i = 1}^{q} \diff X_{i}$,
so that 
\begin{equation}
        G(z) = \lim_{q \to + \infty} \biggl( - \frac{2}{q} \biggr) \partial_{z} \Avg{\ln Z(z)}_{M}. \label{eq:setup-Green}
\end{equation}
We therefore need to compute the expectation value of $\ln Z(z)$ -- essentially a ``quenched average'' \adm{-- which is done using the replica trick}. 
\adm{In particular,} we compute $\avg{Z^{n}(z)}_{M}$ for $n \in \N$ \adm{and analytically continue in $n$} to obtain
\begin{align}\label{eq:setup-continuation}
    \Avg{\ln Z(z)}_{M} = \lim_{n \to 0} \frac{\avg{Z^{n}(z)}_{M} - 1}{n} \, .
\end{align} 
As usual when using the replica trick, we assume the validity of the analytic continuation and replica-symmetry procedure, before taking the thermodynamic limit.

The generalization to other setups is in many cases straightforward. For example, for complex $W \in \C^{p \times q}$, one replaces the transpose by the Hermitian conjugate and modifies Gaussian average and partition function to account for complex-valued integration variables. 
Similarly, more complicated noise spectra \cite{Hachem-2006-Variance, Sengupta-00-Correlations} can be considered by replacing Eq.\,\eqref{eq:Gaussian-noise-moments} with the general Gaussian moments $\langle M_{ia} \rangle_M = \overline{M}_{ia}$ and $\langle M_{ia} M_{jb} \rangle_M = C_{ij} \Sigma_{ab}$. 
In this case, the resulting spectrum can be computed using the same techniques, thereby extending the results of Sengupta and Mitra \cite{Sengupta-99-Distributions} for unperturbed noise, but the resulting coupled equations are more involved and must generally be studied numerically.

Below, we restrict ourselves to real $W = W_0 + M$ with centered i.i.d. noise, and we conventionally formulate our results in terms of the limiting eigenvalue density $P$ of $W^{\top} W$. 
These results can be transferred directly to singular values, which exhibit qualitatively identical behavior.

\section{Outliers for White Noise} 
\adm{We will begin by assuming that the signal matrix $W_0$ has $r$ non-zero singular values $\{ \vartheta_{i} \}_{i = 1, \dotsc, r}$.} 
The derivation follows the approach of Sengupta and Mitra~\cite{Sengupta-99-Distributions} and now includes an explicit added signal matrix. 
Details are found in the Appendix, but can be summarized as follows. 
\adm{The average $\avg{Z^{n}(z)}_{M}$ can be interpreted as the partition function of $n$ copies of the system, and is computed by 
introducing certain auxiliary variables (implementing a Hubbard-Stratonovich transformation).} 
\adm{The resulting integral has a saddle point in the thermodynamic large-$q$ limit.} 
\adm{The corresponding saddle point has replica symmetry, in the sense that it is symmetric under the interchange of copies of the system. 
As a result the equations for the saddle point can be rewritten in a simple form,
as a single equation for the trace of the resolvent, or Green function, $G(z)$:
\begin{align}\label{eq:self-consistency-with-sum}
    G(z) &= \frac{1}{z \bigl[ 1 - \frac{\sigma^2}{\AA} \, G(z) \bigr] - \sigma^2 \bigl( 1 - \frac{1}{\AA} \bigr)} \Biggl( 1 + \lim_{q \to + \infty} \frac{1}{q} \nonumber \\
    &\hspace{-7mm}\times \sum_{i = 1}^{r} \frac{\vartheta^2_{i}}{\bigl[ 1 - \frac{\sigma^2}{\AA} \, G(z) \bigr] \bigl( z [ 1 - \frac{\sigma^2}{\AA} \, G(z)] - \sigma^2 \bigl( 1 - \frac{1}{\AA} \bigr) \bigr) - \vartheta^2_{i}} \! \Biggr)
\end{align}}

To understand this equation, first consider the case where $r$ remains finite in the large-$q$ limit.
Then the second line of Eq.\,\eqref{eq:self-consistency-with-sum} vanishes for generic values of $z$, leaving us with a quadratic equation for $G(z)$ whose solutions are
\begin{align}
    G_{\pm}^{\mathrm{MP}}(z) &= \frac{\AA}{2 \sigma^2 z} \Bigl[ z - \sigma^2 (1 - 1/\AA) \nonumber \\
    &\quad \pm \sqrt{[z - \sigma^2 (1 + 1/\AA)]^2 - 4 \sigma^4 / \AA} \Bigr] .
\end{align}
The solution $G_{-}^{\mathrm{MP}}$ is precisely the resolvent of the Marchenko-Pastur distribution, which reflects the fact that, in the large-$q$ limit, finitely many signal eigenvalues will not alter the noise ``bulk''.
At particular values of $z$, however, the denominator in the second line of Eq.\,\eqref{eq:self-consistency-with-sum} vanishes and $G(z)$ diverges; this indicates the existence of an outlying single eigenvalue that emerges from the bulk.
The locations of these divergences are found by setting 
$G = G_{-}^{\mathrm{MP}}$, which reproduces the usual BBP phase transition \cite{BBP-05-Transition}. 
We conclude that in this case 
the largest $r$ eigenvalues of 
$\tran{W} W$ are given by
\begin{align}\label{eq:MP-BBP-EVs}
    x_{i} = \begin{cases}
        \frac{(\vartheta_{i}^2 + \sigma^2) (\vartheta_{i}^2 + \sigma^2 / \AA)}{\vartheta_{i}^2} \comma &\vartheta_{i}^2 \geq \sigma^2 \mathcal{A}^{-1/2} \\
        \sigma^2 (1 + 1 / \sqrt{\mathcal{A}})^2 \comma &\text{else}
    \end{cases}
\end{align}
in the large $q$ limit. 
This reproduces a special case explored by Benaych-Georges and Nadakuditi~\cite{Benaych-12-Singular}, which here holds for arbitrary low-rank $W_0$. 
Hoyle and Rattray~\cite{Hoyle-03-Outliers, Hoyle-04-PCA} found this relation for a spiked covariance model, as did Baik and Silverstein~\cite{Baik-2006-Spiked} for a Johnstone-spiked population model. 
Thus, the signal-plus-i.i.d.-noise model and the spiked-covariance setup exhibit identical bulk and outlier behavior.

When $r$ scales linearly with $q$, however, the second line of Eq.\,\eqref{eq:self-consistency-with-sum} is order one for generic values of $z$, and the spectrum is no longer Marchenko-Pastur. 
The simplest case is when $W_0$ has an $r$-fold degenerate singular value $\vartheta$, with $\tilde{r} = r/q \in (0, 1)$ fixed and non-zero.
We can then write the self-consistency equation as
\begin{align}\label{eq:self-consistency-MP-and-signal}
    &- \frac{z \sigma^2}{\AA} \bigl[ G(z) - G_{-}^{\mathrm{MP}}(z) \bigr] \bigl[ G(z) - G_{+}^{\mathrm{MP}}(z) \bigr] \nonumber \\
    &= \frac{\tilde{r}\vartheta^2}{\bigl[ 1 - \frac{\sigma^2}{\AA} \, G(z) \bigr] \bigl( z \bigl[ 1 - \frac{\sigma^2}{\AA} \, G(z) \bigr] - \sigma^2 \bigl( 1 - \frac{1}{\AA} \bigr) \bigr) - \vartheta^2} 
\end{align}
separating the influences of noise on the left and signal on the right. 
This is a fourth-order polynomial in $G(z)$ which can be solved exactly; see Appendix~\ref{app:green-function-solution}.
Only one of these four solutions has non-negative density of states, allowing us to determine the spectrum analytically.

Figure~\ref{fig:spectrum-splitting} shows the corresponding density for several signal strengths, compares it to the Marchenko-Pastur law, and confirms it with numerical simulations. 
As the rank ratio $\tilde{r}$ decreases, the bulk part of the spectrum approaches the Marchenko-Pastur law. 
As the signal $\vartheta^2$ increases, the spectrum splits into two parts, with a fraction $1-\tilde{r}$ of the eigenvalues forming 
a bulk similar to Marchenko-Pastur, and the remaining eigenvalues forming a continuous ``signal'' part of the distribution.
The critical signal strength at which this separation occurs marks a phase boundary, and is our generalization of the BBP phase transition to signals of extensive rank.

\begin{figure}[t]
    \centering
    \includegraphics[width=\linewidth, height=5.3cm, keepaspectratio=true]{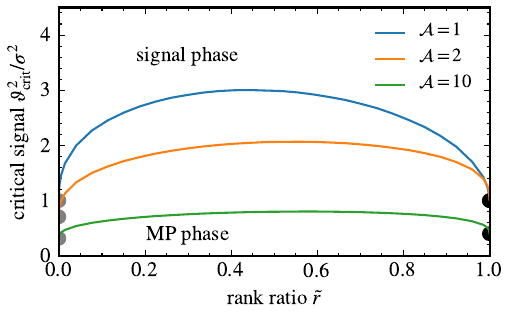} 
    \vspace{0.5mm}
    \caption{Phase diagram for signal and rank ratio, plotted for $\AA \in \{ 1, 2, 10 \}$, $\sigma = 1$. 
    In the Marchenko-Pastur (MP) phase the spectrum has one bulk, in the signal phase two -- the noise and signal bulk. 
    Numerical procedure: for fixed $\tilde{r}$, the signal is increased until the discriminant of Eq.\,\eqref{eq:self-consistency-with-sum}, see Appendix \ref{app:green-function-solution}, no longer gives two but four real and positive roots (or three at the exact phase transition). 
    Each root corresponds to a spectral boundary.
    The bisection method can serve as a faster alternative to find the precise transition point.} 
    \label{fig:phase-diagram}
\end{figure}

This behavior -- the emergence of a signal distribution at a critical value of $\vartheta^2$ -- can be found for every rank ratio $\tilde{r}$. 
As a result, we obtain the phase diagram in Fig.\,\ref{fig:phase-diagram} as $\vartheta^2$ and $\tilde r$ are varied. 
The Marchenko-Pastur phase is characterized by a single bulk, the signal phase by two disconnected parts of the spectrum. 
As $\tilde r\to 0$, the location of the phase transition approaches the expected BBP value $\vartheta^2_{\mathrm{crit}} / \sigma^2 = \AA^{-1/2}$ from Eq.\,\eqref{eq:MP-BBP-EVs}. \par

The full-rank limit $\tilde{r} \to 1$ can be treated similarly to the $\tilde{r} \to 0$ limit. 
Equation~\eqref{eq:self-consistency-with-sum} splits into bulk and outlier contributions, now with the noise-broadened signal producing the bulk and the noise-only part leading to delta peaks. 
A short computation, given in Appendix~\ref{app:full-rank-limit}, leads to the following expression for the $q-r$ smallest eigenvalues:
\begin{align}\label{eq:full-rank-BBP-analog} 
    x_0 &= \begin{cases}
        \sigma^2 \bigl( 1 - \frac{1}{\AA} \bigr) \bigl( 1 - \frac{\sigma^2}{\AA} \frac{1}{\vartheta^2} \bigr) \ , \ &\vartheta^2 \geq \frac{\sigma^2}{\AA} \bigl( 1 + \sqrt{\AA - 1} \bigr) \\
        \sigma^2 \bigl( 1 - \frac{1}{\AA} \bigr) \frac{\sqrt{\AA - 1}}{1 + \sqrt{\AA - 1}} \ , \ &\mathrm{else}.
    \end{cases}
\end{align}
We refer to this as the ``analog BBP'' phase transition, marked by the black dot in Fig.\,\ref{fig:phase-diagram}. 

The phase diagram reveals several features: 
1) For each aspect ratio $\AA$, the critical signal forms an arc between the low-rank and full-rank regime, which mark the corresponding endpoints. 
This shows that the presence of a degenerate signal singular value requires a larger signal strength than in either asymptotic regime. 
2) As the aspect ratio $\AA$ increases and the matrix $W$ is stretched, the arc flattens and the critical signal changes less relative to the BBP point.
3) Both the low-rank and full-rank asymptotic regimes display sharp slopes, indicating a significant crossover between finitely and extensively many outliers.

\begin{figure}[t]
    \centering
    \includegraphics[width=\linewidth]{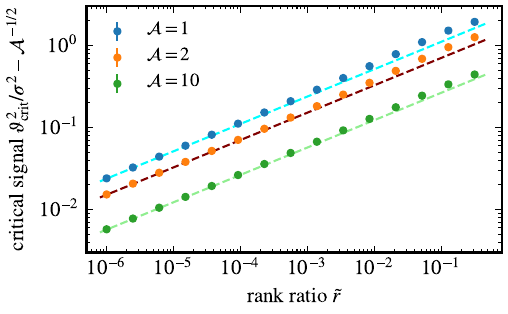}
    \caption{Zoom-in on the low-rank regime. 
    By subtracting $\AA^{-1/2}$ from the normalized critical signals, we reveal the convergence to the BBP phase transition in the limit $\tilde{r} \to 0$ with uniform scaling. 
    The dashed lines show the predicted scalings $\sim \tilde{r}^{1/3}$ according to Eq.\,\eqref{eq:low-rank-scaling}.
    The dots are determined algorithmically (bisection method) using the discriminant of Eq.\,\eqref{eq:self-consistency-with-sum} to find the spectral boundaries; asymptotically, they exhibit perfect agreement.
    The error is set as the last update from the bisection algorithm and of order $10^{-4}$.
    }
    \label{fig:scaling-log}
\end{figure}

To understand the low-rank limit, we analyze the phase transition for small $\tilde{r}$ and find that the critical signal $\vartheta^2_{\mathrm{crit}}$ at the bifurcation scales as $\tilde{r}^{1/3}$.
Interestingly, the exponent is independent of the aspect ratio and appears to be a universal property of the finite-to-extensive-rank crossover.
We validate this scaling numerically for different aspect ratios in Fig.\,\ref{fig:scaling-log}.
Appendix~\ref{app:scaling-proof} gives the derivation in detail, containing an exact solution of the critical signal as well as the spectral boundaries in the case of square matrices.
The main ideas to prove the scaling law are as follows: we write the self-consistency equation as a polynomial and eliminate the Green function $G$ by considering the discriminant; then, we repeat this step to eliminate the eigenvalue variable $z$ through another discriminant.
Subsequently, we employ perturbation theory by expanding around the critical BBP signal in powers of $\tilde{r}^{k}$ for some $k > 0$ and finally determine the smallest exponent, including $k$, using the method of dominant balance.

We outline the derivation here, the lengthy explicit expressions are given in Appendix~\ref{app:scaling-proof}.
We first turn the self-consistency equation for the Green function into a fourth-order polynomial in $G$ and consider its discriminant, whose roots are the spectral boundaries. 
This discriminant factors into the form
\begin{align}
    \Delta(z, \vartheta^2, \tilde{r}) = \frac{\sigma^{36}}{\AA^{13}} \biggl( \frac{z}{\sigma^2} \biggr)^{\! 4} \cdot p^{(5)} \biggl( \frac{z}{\sigma^2}, \frac{\vartheta^2}{\sigma^2}, \tilde{r} \biggr)
\end{align}
where $p^{(5)}$ a fifth-order polynomial in $z / \sigma^2$.
Because the boundaries $z$ cannot be obtained directly in closed form, so we isolate the spectral splitting, i.e. the bifurcation problem, by considering the discriminant of this fifth-order polynomial,
\begin{align}
    \hat{\Delta} \biggl( \frac{\vartheta^2}{\sigma^2}, \tilde{r} \biggr) = \biggl[ f \biggl( \frac{\vartheta^2}{\sigma^2}, \tilde{r} \biggr) \biggr]^3 \cdot p^{(2)} \biggl( \frac{\vartheta^2}{\sigma^2}, \tilde{r} \biggr)
\end{align}
Here, $f$ and $p^{(2)}$ are again polynomials, now in the variable $y := \vartheta^2 / \sigma^2$. 
The function $p^{(2)}$ does not admit relevant roots $\vartheta^2 > 0$.
It remains to find those of $f$, a ninth-order polynomial in $y$.
One of these roots the physical solution for the critical signal as a function of the rank ratio.
The elimination of variables via discriminants comes at the cost of increasing the polynomial degree in the remaining variables, thence a closed-form solution is not available in general.
Employing perturbation theory, we first solve the unperturbed problem $\tilde{r} = 0$,
\begin{align}\label{eq:unperturbed-function}
    f(y_0, 0) &= \AA (1 + \AA + 2 \AA y_0)^3 (-1 + \AA y_0^2)^3 \overset{!}{=} 0 \ ,
\end{align} 
yielding the usual BBP critical signal $\vartheta^2_0 = \sigma^2 y_0 = \sigma^2 \mathcal{A}^{-1/2}$ from the latter bracket.
A two-variable Taylor expansion around the point $(y_0, 0)$ with $y = y_0 + C_1 \tilde{r}^{k} + C_2 \tilde{r}^{2k} + \mathcal{O}(\tilde{r}^{3k})$ for some $k > 0$ gives
\begin{align}
    f(y, \tilde{r}) 
    &= \frac{\partial f}{\partial \tilde{r}} (y_0, 0) \cdot \tilde{r} + \frac{1}{3!} \frac{\partial^3 f}{\partial^3 y} (y_0, 0) \cdot C_1^3 \tilde{r}^{3 k} 
    \nonumber \\
    &\quad + \mathcal{O}(\tilde{r}^{1+k}) + \mathcal{O}(\tilde{r}^2)
\end{align}

\begin{figure}[t]
    \includegraphics[width=0.99\linewidth]{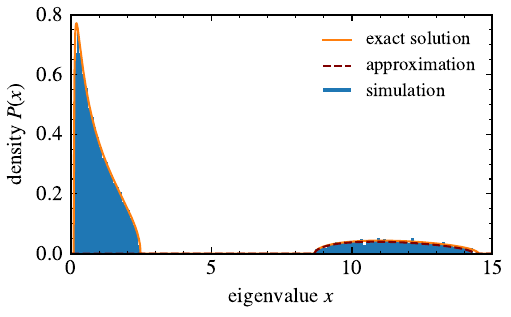}
    \caption{Approximate spectral density by Eq.\,\eqref{eq:approximate-outlier-PDF} in comparison with the exact spectrum and simulations for a $q = 500$ matrix averaged over 10 runs. $\AA = 2$, $\sigma = 1$, $\tilde{r} = 0.2$, $\vartheta^2 = 10$.}
    \label{fig:approx-spectrum}
\end{figure}

\noindent
The coefficients of $\tilde{r}^{k}$ and $\tilde{r}^{2k}$ are identically zero here, because the solution for $\tilde{r} = 0$ of Eq.\,\eqref{eq:unperturbed-function} stems from a triple root. 
Thus, the BBP solution bifurcates into three roots and we can expect the first-order perturbation term to resemble that.
By the method of dominant balance \cite{Bender-1999-Perturbation}, the lowest-order terms, here $\tilde{r}$ and $\tilde{r}^{3k}$, must be of equal order such that their combined contribution can vanish.
This not only gives $k = 1/3$ but also specifies $C_1$ after reinserting $k$ into the above expansion.
Finally, we have $y = y_0 + C_1 \tilde{r}^{1/3} + \mathcal{O}(\tilde{r}^{2/3})$ and, with the explicit value of $C_1$,
\begin{align}\label{eq:low-rank-scaling}
    \frac{\vartheta^2_{\mathrm{crit}}}{\sigma^2} - \frac{1}{\sqrt{\AA}}
    &= \frac{3}{2} \frac{(1 + \sqrt{\mathcal{A}})^{2/3}}{\mathcal{A}^{5/6}} \cdot \tilde{r}^{1/3} + \mathcal{O}(\tilde{r}^{2/3})
\end{align}
This power law explains the sharp increase in the critical signal when the degeneracy of a signal eigenvalue stops being finite and starts becoming extensive.
The effect of the noise-induced broadening of the outlier becomes highly significant, resulting in a substantial increase of the signal strength required in order to produce separate signal and noise bulks.

\begin{figure}[t]
    \includegraphics[width=\linewidth]{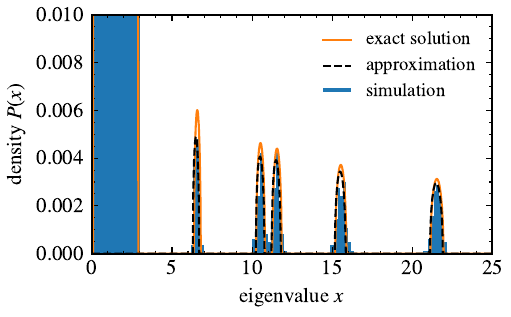}
    \caption{Spectral density for five distinct signals $\{ \vartheta^2_{i} \}_{i = 1, \dotsc, 5} = \{ 5, 9, 10, 14, 20 \}$, each degenerate with a rank ratio $\tilde{r}_{i} = 1 / 500$. 
    The exact solution is determined numerically via fixed-point iteration of Eq.\,\eqref{eq:self-consistency-arbitrary-signal}, as an approximate solution we add up the asymptotic density \eqref{eq:approximate-outlier-PDF} for each signal separately, and the histogram stems from numerical simulations. 
    We conduct 100 runs for a finite $p \times q$ dim. matrix with $q = 500$, $p = \AA q$, and $\AA = 2$. 
    Notably, the broadening increases with larger signals.}
    \label{fig:multiple-signals}
\end{figure}

\section{Large Signals} 
In many applications the signal is large compared to the noise, and it is convenient to describe the two components of our distribution -- the noise bulk and signal bulk -- independently using an asymptotic approximation to Eq.\,\eqref{eq:self-consistency-MP-and-signal}. 
In particular, when $\vartheta^2 \gg \sigma^2$ we expect the signal to be concentrated around values of $z = \vartheta^2 +{\cal O}(1)$ with $|z| \gg \sigma^2$.  
Expanding~\eqref{eq:self-consistency-MP-and-signal} in powers of $1/z$, 
and using
$G_{-}^{\mathrm{MP}}(z) = z^{-1} + \mathcal{O}(z^{-2})$ and $G_{+}^{\mathrm{MP}}(z) = \AA \sigma^{-2} - \AA z^{-1} + \mathcal{O}(z^{-2})$, gives
\begin{align}
    0 &= z^2 G^2(z) \biggl[ - \frac{2 \sigma^2}{\AA} - \frac{\sigma^2}{\AA} \frac{z - \vartheta^2}{z} \biggr] \nonumber \\ 
    &\quad + z G(z) \biggl[ \alpha(z) - \sigma^2 (1 - 1/\AA) \frac{z - \vartheta^2}{z} \biggr] \nonumber \\
    &\quad- \biggl[ \alpha(z) + \tilde{r} \vartheta^2 - \frac{2 \sigma^2}{\AA} - \sigma^2 \frac{z - \vartheta^2}{z} \biggr] + \mathcal{O}(z^{-1}) 
\end{align}
where $\alpha(z) := z - \vartheta^2 - \sigma^2 (1 - 3/\AA)$.  
Neglecting terms of order $\mathcal{O}(z^{-1})$, including $(z - \vartheta^2) / z$ under the above assumption, yields a quadratic equation for $G$:
\begin{align}
    0 &= - \frac{2 \sigma^2}{\AA} \, z^2 G^2(z) + \alpha(z) z G(z) - \biggl[ \alpha(z) + \tilde{r} \vartheta^2 - \frac{2 \sigma^2}{\AA} \biggr] 
\end{align}
The two solutions are 
\begin{align}\label{eq:approx-Green-large-signal}
    G_{\mathrm{aprx}, \pm}(z) &= \frac{\alpha(z) \pm \sqrt{\alpha^2(z) - \frac{8 \sigma^2}{\AA} \bigl( \alpha(z) + \tilde{r} \vartheta^2 - \frac{2 \sigma^2}{\AA} \bigr)}}{4 z \sigma^2 / \AA} 
\end{align}
where $G_{\mathrm{aprx}, -}$ 
corresponds to the non-negative eigenvalue density 
\begin{align}\label{eq:approximate-outlier-PDF}
    P_{\mathrm{aprx}}(x) &= \frac{\AA}{4 \pi x \sigma^2} \biggl\{ - [x - \vartheta^2 - \sigma^2 (1 - 3/\AA)]^2 \nonumber \\
    &\quad + \frac{8 \sigma^2}{\AA} \Bigl[ x - (1 - \tilde{r}) \vartheta^2 - \sigma^2 (1 - 1/\AA) \Bigr] \biggr\}^{1/2} \ .
\end{align}
This formula approximates the outlier bulk, i.e., the signal part, and is valid in the large-$\vartheta$ limit. 
This expression, somewhat similar to the Marchenko-Pastur law, demonstrates that even in the large signal limit the noise term effectively broadens the distribution of signal eigenvalues. 
The accuracy of this approximation at large signal is demonstrated in
Fig.\,\ref{fig:approx-spectrum}, where it is compared both to  
the exact solution from the fourth-order polynomial and to numerical simulations.

\section{Arbitrary Signal Matrices} 
Most applications involve not a single degenerate signal but many distinct signals. 
In this case solving Eq.\,\eqref{eq:self-consistency-with-sum} becomes more involved.
For example, if the signal matrix has $n$ distinct eigenvalues, each with an eigenspace whose rank scales linearly with $q$, then Eq.\,\eqref{eq:self-consistency-with-sum} is a polynomial equation of order $2n + 2$. 
This also becomes clear when expressing the problem via the Green function's inverse~\cite{Silverstein-95-Existence}.
More generally, for arbitrary signal matrices we have
\begin{align}\label{eq:self-consistency-arbitrary-signal}
    G(z) &= \int_{0}^{+ \infty} \frac{\rho_0 (\vartheta^2) \ \diff (\vartheta^2)}{z \bigl[ 1 - \frac{\sigma^2}{\AA} \, G(z) \bigr] - \sigma^2 \bigl( 1 - \frac{1}{\AA} \bigr) - \frac{\vartheta^2}{1 - \frac{\sigma^2}{\AA} \, G(z)}} 
\end{align}
where $\rho_0$ is the eigenvalue density of $\tran{W_0} W_0$. 
This relation reproduces the result of Dozier and Silverstein~\cite{Dozier-07-Green}, where $g(z) = (1 - 1/\AA) / z + G(z) / \AA$ is used as the Green function (Stieltjes transform, or trace of the resolvent of $W W^{\top}$), $\varrho_0(\vartheta^2) = (1 - 1/\AA) \delta(\vartheta^2) + \rho_0(\vartheta^2) / \AA$ as a signal eigenvalue density, and $\sigma^2 / q = \AA \sigma^2 / p$ instead of $\sigma^2 / p$ in the second moments~\eqref{eq:Gaussian-noise-moments}. 
The replica derivation thereby gives a complementary route to established identities for this problem. 
\par

\begin{figure}[t]
    \includegraphics[width=\linewidth]{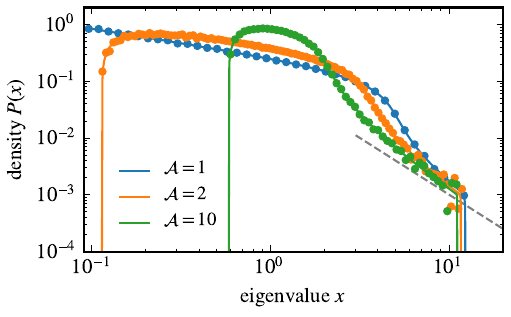}
    \caption{Theoretical predictions (lines) via Eq.\,\eqref{eq:self-consistency-power-law-signals} and numerical simulations (dots, instead of histograms) with signal eigenvalues distributed according to an inverse square power law between $k = 0.1$ and $K = 10$. 
    We conduct 100 runs for a finite $p \times q$ dim. matrix with $q = 1000$, $p = \AA q$. 
    The dashed line is proportional to $x^{-2}$.
    Each spectrum assumes the power-law structure in the upper part of the spectrum.}
    \label{fig:power-law-signals}
\end{figure}

In many applications, such as neural networks, the signals may constitute a non-negligible portion of the total rank \cite{Thamm-2022-RMT}.
In the large-$q$ limit, these signals may either cluster around a finite number of values, yielding a finite-order polynomial equation for $G(z)$, or form a continuous distribution, requiring the more general integral equation.
Figure \ref{fig:multiple-signals} shows an example with five distinct signal eigenvalues, each with non-zero rank ratio. 
As before, the exact solution, approximation, and numerical simulations agree well. 
The exact solution is determined via Eq.\,\eqref{eq:self-consistency-arbitrary-signal}, where $\rho_0$ is a sum of equally weighted delta peaks. 
This gives an order-12 polynomial, which can be written as a fixed-point equation and solved iteratively.   
Adjacent signal bulks are well approximated by treating them as separate large-signal components, as in the previous section.

Another example, perhaps more useful for applications, is a full-rank signal matrix whose singular value distribution exhibits a power law \cite{Maloney-22-Network}. 
Assuming $\rho_0(x) \sim x^{-2}$ on the interval $(k, K) \subset \R$, the integral in Eq.\,\eqref{eq:self-consistency-arbitrary-signal} can be evaluated exactly, giving 
\begin{align}\label{eq:self-consistency-power-law-signals}
   G(z) &= \Biggl[ 1 + \frac{1}{\frac{1}{k} - \frac{1}{K}} \frac{1}{C(z, G)} \log \Biggl( \frac{1 - \frac{C(z, G)}{k}}{1 - \frac{C(z, G)}{K}} \Biggr) \Biggr] \nonumber \\
   &\quad \times \frac{1 - \frac{\sigma^2}{\AA} \, G(z)}{C(z, G)} 
\end{align}
with $C(z, G) = \bigl[ 1 - \frac{\sigma^2}{\AA} \, G(z) \bigr] \bigl( z \bigl[ 1 - \frac{\sigma^2}{\AA} \, G(z) \bigr] - \sigma^2 (1 - \frac{1}{\AA}) \bigr)$.
The Green function can be determined numerically from this self-consistency equation.
Figure~\ref{fig:power-law-signals} depicts the predicted eigenvalue densities and numerical simulations, which again show excellent agreement.
\par

\section{Discussion} 
In the work at hand, we investigated the BBP-type phase transition for an extensive number of signals in a signal-plus-noise matrix model. 
Utilizing methods from Random Matrix Theory, in particular the replica trick, we developed a framework to explore the transition from low-rank signals to extensive-rank signals. 
We demonstrated the splitting of the spectral density into two distinct bulks, mapped the associated phase diagram, derived a new scaling law in the finite-to-extensive-rank crossover, and showed that a large-signal approximation accurately captures the density's outlier part -- even for multiple distinct signals -- with a further example for a power-law signal distribution in the more general setting. \par 

These results extend the BBP picture of a low-rank signal beyond isolated outliers towards finite-density signal bulks, thereby addressing regimes that are relevant to a broad range of applications.
In particular, the $1/3$-scaling highlights that the passage from finitely many to extensively many outliers is singular: even a small but non-vanishing rank ratio requires a parametrically larger signal to separate from the noise bulk.
The resulting formulas, especially the large-signal approximation, give a compact analytic description of the broadened signal spectrum and may serve as useful benchmarks for numerical and inference-based studies.
\par




\section*{Acknowledgments}
The authors gratefully acknowledge useful conversations with B. M{\' e}nard.
This work was performed in part at the Aspen Center for Physics, which is supported by National Science Foundation grant PHY-2210452. The work of A.M.~and B.R.~is supported in part by the Simons Foundation Grant No.~12574. 
Research of A.M. is supported in part by the Natural Sciences and Engineering Research Council of Canada (NSERC), funding reference number SAPIN/00047-2020. B.R. would like to acknowledge support by DFG grant RO/2247/16-1.

\bibliography{bib}

\appendix

\vspace{2em}

\section{Details on the Replica Derivation}\label{app:replica} 
In the quantity $Z^{n}(z)$, we introduce new integration variables via inclusion of unity through Gaussian integrals. With a change of variables we get rid of the quadratic terms in $W$. Afterwards, we take the average w.r.t. $M$ and conduct a Hubbard-Stratonovich transformation, 
\begin{align}\label{eq:general-partition-avg-after-trafo} 
    & \avg{Z^{n}(z)}_{M} = \biggl( \frac{q}{2 \pi} \biggr)^{n (p + q)/2} \int_{\R^{n (p+q)}} \prod_{\alpha = 1}^{n} \Diff X_{\alpha} \, \Diff Y_{\alpha} \nonumber \\
    &\quad \times \ee^{- \frac{q}{2} \sum_{\alpha = 1}^{n} (\tran{X_{\alpha}} z X_{\alpha} + \tran{Y_{\alpha}} Y_{\alpha} + \tran{Y_{\alpha}} W_0 X_{\alpha} + \tran{(W_0 X_{\alpha})} Y_{\alpha})} \nonumber \\
    &\quad \times \ee^{\frac{q}{2} \frac{\sigma^2}{\AA} \Tr \sum_{\alpha, \beta = 1}^{n} X_{\alpha} \tran{Y_{\alpha}} Y_{\beta} \tran{X_{\beta}}} \cdot \frac{1}{(2 \pi \sigma^2 / p)^{pq/2}} \nonumber \\
    &\quad \times \int_{\R^{pq}} \Diff M \ \exp \biggl\{- \frac{p}{2 \sigma^2} \Tr \biggl[ \biggl( M + \frac{\sigma^2}{\AA} \sum_{\alpha = 1}^{n} Y_{\alpha} \tran{X_{\alpha}} \biggr)^{\top} \nonumber \\
    &\quad \times  \biggl( M + \frac{\sigma^2}{\AA} \sum_{\alpha = 1}^{n} Y_{\alpha} \tran{X_{\alpha}} \biggr) \biggr] \biggr\} \ .
\end{align}
The average can be readily performed. Regarding the quartic terms $\tran{X_{\beta}} X_{\alpha} \tran{Y_{\alpha}} Y_{\beta}$, we insert unity through a product of delta distributions in their integral representation. Herein, we replace $\tran{X_{\beta}} X_{\alpha}$ by $Q_{\alpha \beta}$ for every $\alpha, \beta \in \{ 1, \dotsc, n\}$. The previous variables are integrated away via a combined high-dimensional Gaussian integral with a matrix $A$ in its quadratic form:
\begin{align}\label{eq:avg-before-SP-approx}
    \avg{Z^{n}(z)}_{M} &= \biggl( \frac{q}{4 \pi} \biggr)^{n^2} \hspace{-1mm} \int_{\R^{2 n^2}} \Diff Q \, \Diff R \ \ee^{- \frac{q}{2} [\Tr (\ii R Q) + \frac{1}{q} \Tr \ln A_{+}]}
\end{align}
where $A_{+}$ is the symmetrized \cite{Mondaini-17-Gaussian} version $A_{+} := (A + \tran{A}) / 2$ of $A$, which is defined below. The Gaussian integral could be evaluated because singular matrices have Lebesgue measure zero, i.e. $\det A_{+} \neq 0$ almost surely \cite{Brock-05-Scarcity}. It is convenient to label the matrix $A$ using the indices $\alpha, \beta$, therewith it is defined as
\begin{align}\label{eq:def-matrix-A-blocks} 
    A_{\alpha \beta} := \begin{pmatrix}
        (z \delta_{\alpha \beta} - \ii R_{\alpha \beta}) \oneq & \delta_{\alpha \beta} \tran{W_0} \\
        \delta_{\alpha \beta} W_0 & \bigl( \delta_{\alpha \beta} - \frac{\sigma^2}{\AA} \, Q_{\alpha \beta} \bigr) \onep
    \end{pmatrix} \ .
\end{align}
Observe how each entry is a matrix of dimension $p + q$ itself so that the exponent in Eq.\,\eqref{eq:avg-before-SP-approx} indeed scales with $q$ in total. Now, we apply a saddle-point approximation for Eq.\,\eqref{eq:avg-before-SP-approx} and choose a replica-diagonal ansatz letting $Q_{\alpha \beta} = \delta_{\alpha \beta} \QQ(z)$ and $R_{\alpha \beta} = - \ii \delta_{\alpha \beta} \RR(z)$. Only the highest-order term $\avg{Z^{n}(z)}_{M} \sim \ee^{- \frac{nq}{2} \SS(z, \QQ(z), \RR(z))}$ survives the analytic continuation. The maximizing exponent contains the self-energy 
\begin{align}\label{eq:SP-exponent-general-form-2} 
    &\SS(z, \QQ(z), \RR(z)) = \RR(z) \QQ(z) + \AA \ln \biggl( 1 - \frac{\sigma^2}{\AA} \, \QQ(z) \biggr) \nonumber \\
    &\qquad + \frac{1}{q} \Tr \ln \biggl( [z - \RR(z)] \oneq - \frac{\tran{W_0} W_0}{\onep - \frac{\sigma^2}{\AA} \, \QQ(z)} \biggr) 
\end{align}
which uses the Schur complement to rewrite the determinant of a matrix such as $A_{\alpha \beta}$. The saddle-point equations are given by $\partial_{\QQ} \SS = 0$, $\partial_{\RR} \SS = 0$. They are quite similar and allow for a simple relation between $\QQ(z)$ and $\RR(z)$. Then, a full description is given by
\begin{eqnarray}\label{eq:SP-equations-MP-noise}
        \QQ(z) = \frac{1}{z - \RR(z)} \Biggl( \! 1 + \frac{1}{q} 
        \sum_{i = 1}^{r} \frac{\vartheta_{i}^2}{\bigl[ 1 - \frac{\sigma^2}{\AA} \QQ(z) \bigr] [z - \RR(z)] - \vartheta_{i}^2} \! \Biggr)
        \nonumber\\
        \label{eq:SP-equations-MP-noise-2}
\end{eqnarray}
and 
\begin{align}\label{eq:relation-Q-R}
    \frac{\RR(z)}{z \sigma^2} = \frac{1 - 1/\AA}{z} + \frac{\QQ(z)}{\AA} \ .
\end{align}
Moreover, observing $\partial_{R} \SS = \QQ(z) - \partial_{z} \SS$, the Green function is found to be
\begin{align}\label{eq:Green-limit-via-S}
    G(z) &= \lim_{q \to + \infty} \partial_{z} \SS(z, \QQ(z), \RR(z)) 
\end{align}
such that $\QQ(z)$ is related to the Green function via $G(z) = \lim_{q \to + \infty} \QQ(z)$.

\section{Details on the Analytic Green Function Solution}
\label{app:green-function-solution}
Eq.\,\eqref{eq:self-consistency-MP-and-signal} is a fourth-order polynomial with a few algebraic manipulations: $0 = A(z) G^4(z) + B(z) G^3(z) + C(z) G^2(z) + D(z) G(z) + E(z)$ where
\begin{align}\label{eq:Green-polynomial}
    A(z) &= z^2 \frac{\sigma^6}{\AA^3} 
    \\
    B(z) &= \frac{\sigma^4}{\AA^2} \biggl[ 2 z \sigma^2 \biggl( 1 - \frac{1}{\AA} \biggr) - 3 z^2 \biggr] 
    \\
    C(z) &= \frac{\sigma^2}{\AA} \biggl[ 3 z^2 - z \vartheta^2 + \sigma^4 \biggl( 1 - \frac{1}{\AA} \biggr)^2 \nonumber \\
    &\quad - 4 z \sigma^2 \biggl(1 - \frac{1}{\AA} \biggr) + z \frac{\sigma^2}{\AA}  \biggr] 
    \\
    D(z) &= \biggl[ z \vartheta^2 - \vartheta^2 \sigma^2 \biggl(1 - \frac{1}{\AA} \biggr) + 2 z \sigma^2 \biggl( 1 - \frac{1}{\AA} \biggr)  \nonumber \\
    &\quad - \sigma^4 \biggl( 1 - \frac{1}{\AA} \biggr)^2 - z^2 + \frac{\sigma^4}{\AA} \biggl( 1 - \frac{1}{\AA} \biggr) - 2 z \frac{\sigma^2}{\AA} \biggr] 
    \\
    E(z) &= z - \sigma^2 \biggl( 1 - \frac{1}{\AA} \biggr) - (1 - \tilde{r}) \vartheta^2 \ . 
\end{align}
This representation is only worthwhile for non-zero $\tilde{r}$. 
The full solution reads:
\begin{align}\label{eq:Green-poly-solutions}
    G_{1, \dotsc, 4}(z) &= - \frac{B(z)}{4 A(z)} \mp_{(I)} \frac{1}{2} \sqrt{2 y - a} \nonumber \\
    &\quad \pm_{(II)} \frac{1}{2} \sqrt{- 2 y - a \pm_{(I)} \frac{2 b}{\sqrt{2 y - a}}}
\end{align}
where $\pm_{(I)}$ and $\pm_{(II)}$ are unrelated signs so as to give four sign combinations in total, 
\begin{subequations}
   \begin{align}
        y &= y(z) := \frac{a}{6} + \sqrt[3]{- \frac{n}{2} + \sqrt{\frac{n^2}{4} + \frac{m^3}{27}}} \nonumber \\
        &\qquad\qquad - \frac{m}{3} \frac{1}{\sqrt[3]{- \frac{n}{2} + \sqrt{\frac{n^2}{4} + \frac{m^3}{27}}}} \\
        m &= m(z) := - \biggl( \frac{a^2(z)}{12} + c(z) \biggr) \ , \\
        n &= n(z) := - \frac{a^3(z)}{108} + \frac{a(z) c(z)}{3} - \frac{b^2(z)}{8}
    \end{align}
\end{subequations}
and
\begin{subequations}
    \begin{align}
        a(z) &:= - \frac{3 B^2(z)}{8 A^2(z)} + \frac{C(z)}{A(z)} \ , \\
        b(z) &:= \frac{B^3(z)}{8 A^3(z)} - \frac{B(z) C(z)}{2 A^2(z)} + \frac{D(z)}{A(z)} \\
        c(z) &:= - \frac{3 B^4(z)}{256 A^4(z)} + \frac{C(z) B^2(z)}{16 A^3(z)} - \frac{B(z) D(z)}{4 A^2(z)} + \frac{E(z)}{A(z)} \ .
    \end{align}
\end{subequations}
The polynomial gives rise to the discriminant $\Delta = \Delta(z)$ defined by (\cite{Irving-04-Discriminant})
\begin{align}\label{eq:discriminant}
    \Delta &:= 256 A^3(z) E^3(z) - 192 A^2(z) B(z) D(z) E^2(z) \nonumber \\
    & - 128 A^2(z) C^2(z) E^2(z) + 144 A^2(z) C(z) D^2(z) E(z) \nonumber \\
    & - 27 A^2(z) D^4(z) + 144 A(z) B^2(z) C(z) E^2(z) \nonumber \\
    & - 6 A(z) B^2(z) D^2(z) E(z) \! - \! 80 A(z) B(z) C^2(z) D(z) E(z) \nonumber \\
    & + 18 A(z) B(z) C(z) D^3(z) + 16 A(z) C^4(z) E(z) \nonumber \\
    & - 4 A(z) C^3(z) D^2(z) - 27 B^4(z) E^2(z) \nonumber \\
    & + 18 B^3(z) C(z) D(z) E(z) - 4 B^3(z) D^3(z) \nonumber \\
    & - 4 B^2(z) C^3(z) E(z) + B^2(z) C^2(z) D^2(z) \ .
\end{align}
The discriminant's roots are responsible for the spectrum's boundaries, because they mark a change in the number of real and imaginary solutions. We can furthermore relate the discriminant to one radicand in our full Green function expression:
\begin{align}
    \frac{n^2(z)}{4} + \frac{m^3(z)}{27} = - \frac{27}{4 \cdot [6 A(z)]^6} \, \Delta(z) \ .
\end{align}
The left-hand side is more suitable for the numerical determination of roots due to its smaller scale, leading to fewer fluctuations near zero.

\section{Scaling Law for the Generalized BBP Phase Transition in the Low-Rank Regime}
\label{app:scaling-proof}

\paragraph{Perturbative Approach.}
The phase transition occurs at the point where two of the probability density's roots coincide. 
This happens if the discriminant $\Delta$ from Eq.\,\eqref{eq:discriminant} vanishes (spectral boundaries) \textit{and} when two of its roots are equal (bifurcation).
With the corresponding functions $A(z)$ through $E(z)$, the discriminant $\Delta$ is a polynomial of order $9$ in the eigenvalue variable $z$ ($z = x + \ii \varepsilon \to x$ for $\varepsilon \to 0$).
However, the coefficients for the first four orders are zero, and so a quintic polynomial, which specifies the non-zero solutions, remains.
The discriminant reads
\begin{align}
    \Delta(z) = \frac{\sigma^{36}}{\AA^{13}} \biggl( \frac{z}{\sigma^2} \biggr)^{4} \cdot p^{(5)} \! \biggl( \frac{z}{\sigma^2}, \frac{\vartheta^2}{\sigma^2}, \tilde{r} \biggr)
\end{align}
and the spectral boundaries are given by $\Delta(z) = 0$, implying $0 = p^{(5)} \bigl( \frac{z}{\sigma^2}, \frac{\vartheta^2}{\sigma^2}, \tilde{r} \bigr)$. 
Let us use an easier variable $y := \vartheta^2 / \sigma^2$.
Then, the function $p^{(5)}$ reads
\begingroup
\allowdisplaybreaks
\begin{align}
    p^{(5)} \! & \biggl( \frac{z}{\sigma^2}, y, \tilde{r} \biggr) 
    = 4 \AA^7 y^3 \biggl( \frac{z}{\sigma^2} \biggr)^5 + \AA^5 y^2 \Bigl[ 1 + 9 (2 - 3 r) r 
    \nonumber \\
    & + 2 \AA (-5 + 9 r + 2 (-4 + 3 r) y) + \AA^2 (1 - 8 y (2 + y)) \Bigr]  
    \nonumber \\
    &\times \biggl( \frac{z}{\sigma^2} \biggr)^4 + 2 \AA^3 y \Bigl[ 2 r + \AA (-1 + 4 \AA - \AA^2 - 3 r 
    \nonumber \\
    & - 3 \AA r + 2 \AA^2 r - 2 (1 + \AA (-7 + (-7 + \AA) \AA) + 13 r 
    \nonumber \\
    & + \AA (23 + 13 \AA) r - 18 (1 + \AA) r^2) y + \AA (11 (1 + \AA)^2 
    \nonumber \\
    & - 31 (1 + \AA) r + 24 r^2) y^2 + 6 \AA^2 (2 + 2 \AA - 5 r) y^3 
    \nonumber \\
    & + 2 \AA^3 y^4) \Bigr] \biggl( \frac{z}{\sigma^2} \biggr)^3
    + \AA^2 \Bigl[ (-1 + \AA)^2 \AA - 2 \Bigl( 6 r 
    \nonumber \\
    & + \AA \Bigl[-3 - 5 r + \AA \bigl( 7 - 10 r + \AA [7 - 5 r 
    \nonumber \\
    & + \AA (-3 + 6 r)] \bigr) \Bigr] \Bigr) y + 2 \AA (3 + \AA (-13 + \AA (-12 
    \nonumber \\
    & + \AA (-13 + 3 \AA))) + 21 r + 3 \AA (9 + \AA (9 + 7 \AA)) r 
    \nonumber \\
    & - 4 (7 + \AA (2 + 7 \AA)) r^2) y^2 - 2 \AA^2 (5 + 5 \AA^3 
    \nonumber \\
    & + \AA^2 (19 - 37 r) + \AA (-1 + 2 r) (-19 + 64 r) 
    \nonumber \\
    & + r (-37 - 128 (-1 + r) r)) y^3 - \AA^3 (23 + 18 \AA 
    \nonumber \\
    & + 23 \AA^2 - 96 (1 + \AA) r + 128 r^2) y^4 
    \nonumber \\
    & - 8 \AA^4 (1 + \AA - 2 r) y^5 \Bigr] \biggl( \frac{z}{\sigma^2} \biggr)^2 + 2 (-1 + \AA)^2 \AA 
    \nonumber \\
    &\times \Bigl[ -\AA (1 + \AA) + (-3 (\AA + \AA^3) + 2 (3 + \AA + \AA^2 
    \nonumber \\
    & + 3 \AA^3) r) y - 2 \AA (1 + \AA^3 + 8 \AA (-1 + r) r + 8 r^2) y^2 
    \nonumber \\
    & - \AA^2 (1 + \AA^2 + 8 (1 - 4 r) r + \AA (-6 + 8 r)) y^3 
    \nonumber \\
    & + \AA^3 (3 + 3 \AA - 16 r) y^4 + 2 \AA^4 y^5 \Bigr] \frac{z}{\sigma^2} 
    \nonumber \\
    & + \Bigl[ (-1 + \AA)^4 (-4 r y + \AA (1 + y)^2) 
    \nonumber \\
    &\times (1 + \AA y (2 - 4 r + \AA y)) \Bigr]
\end{align}
\endgroup
At some critical signal $\vartheta^2_{\mathrm{crit}}$, the bulk splits -- a bifurcation occurs.
We use the discriminant of $p^{(5)}$ to identify that signal, where we find
\begin{align} 
    \hat{\Delta} = \biggl[ f \biggl( \frac{\vartheta^2_{\mathrm{crit}}}{\sigma^2}, \tilde{r} \biggr) \biggr]^3 \cdot p^{(2)} \! \biggl( \frac{\vartheta^2_{\mathrm{crit}}}{\sigma^2}, \tilde{r} \biggr) 
    \overset{!}{=} 0 \ .
\end{align}
The function $p^{(2)}$ does not admit relevant roots $\vartheta^2_{\mathrm{crit}} > 0$, because the only solutions of 
\begin{align} 
    0 &= p^{(2)}(y, \tilde{r})
    = -256 (-1 + \AA)^4 \AA^24 (\AA - \tilde{r}) (-1 + \tilde{r}) \nonumber \\
    &\quad \times \tilde{r} y^6 \Bigl[ 1 + (\AA + 4 \AA y)^2 - 
    2 \AA (1 + (-4 + 8 \tilde{r}) y) \Bigr]^2
\end{align}
are non-positive.
The function $f$ is a polynomial of order 9 in $y = \vartheta^2_{\mathrm{crit}} / \sigma^2$ and given by 
\begingroup
\allowdisplaybreaks
\begin{align}
    f(y, \tilde{r}) &= 
    + 8 \AA^7 y^9 + \Bigl[ 12 \AA^6 + 12 \AA^7 - 87 \AA^6 \tilde{r} \Bigr] y^8
    \nonumber \\
    &\quad + \Bigl[ 6 \AA^5 - 12 \AA^6 + 6 \AA^7 - 84 \AA^5 \tilde{r} - 84 \AA^6 \tilde{r} 
    \nonumber \\
    &\quad+ 384 \AA^5 \tilde{r}^2 \Bigr] y^7 + \Bigl[ \AA^4 - 33 \AA^5 - 33 \AA^6 + \AA^7 
    \nonumber \\
    &\quad - 39 \AA^4 \tilde{r} - 246 \AA^5 \tilde{r} - 39 \AA^6 \tilde{r} + 672 \AA^4 \tilde{r}^2 
    \nonumber \\
    &\quad + 672 \AA^5 \tilde{r}^2 - 1376 \AA^4 \tilde{r}^3 \Bigr] y^6 + \Bigl[ -18 \AA^4 - 12 \AA^5 
    \nonumber \\
    &\quad - 18 \AA^6 - 6 \AA^3 \tilde{r} - 678 \AA^4 \tilde{r} - 678 \AA^5 \tilde{r} - 6 \AA^6 \tilde{r} 
    \nonumber \\
    &\quad + 966 \AA^3 \tilde{r}^2 + 4212 \AA^4 \tilde{r}^2 + 966 \AA^5 \tilde{r}^2 
    \nonumber \\
    &\quad - 4416 \AA^3 \tilde{r}^3 - 4416 \AA^4 \tilde{r}^3 + 4608 \AA^3 \tilde{r}^4 \Bigr] y^5  
    \nonumber \\
    &\quad + \Bigl[-3 \AA^3 + 27 \AA^4 + 27 \AA^5 - 3 \AA^6 - 561 \AA^3 \tilde{r} 
    \nonumber \\
    &\quad - 1416 \AA^4 \tilde{r} - 561 \AA^5 \tilde{r} + 4908 \AA^3 \tilde{r}^2 + 4908 \AA^4 \tilde{r}^2 
    \nonumber \\
    &\quad + 660 \AA^5 \tilde{r}^2 - 12288 \AA^3 \tilde{r}^3 - 4128 \AA^4 \tilde{r}^3 
    \nonumber \\
    &\quad + 9216 \AA^3 \tilde{r}^4 - 12 \AA^2 \tilde{r}^2 \Bigl( -55 
    \nonumber \\
    &\quad + 8 \tilde{r} \bigl[ 43 + 24 \tilde{r} (-4 + 3 \tilde{r}) \bigr] \Bigr) \Bigr] y^4
    \nonumber \\
    &\quad + \Bigl[ 18 \AA^3 + 28 \AA^4 + 18 \AA^5 - 768 \AA^3 \tilde{r} - 768 \AA^4 \tilde{r} 
    \nonumber \\
    &\quad - 204 \AA^5 \tilde{r} + 2856 \AA^3 \tilde{r}^2 + 2004 \AA^4 \tilde{r}^2 + 216 \AA^5 \tilde{r}^2 
    \nonumber \\
    &\quad - 3408 \AA^3 \tilde{r}^3 - 1520 \AA^4 \tilde{r}^3 + 2016 \AA^3 \tilde{r}^4 
    \nonumber \\
    &\quad + 8 \AA \tilde{r}^2 (27 + 2 \tilde{r} (-95 + 126 \tilde{r})) 
    \nonumber \\
    &\quad - 12 \AA^2 \tilde{r} (17 + \tilde{r} (-167 + 4 (71 - 12 \tilde{r}) \tilde{r})) \Bigr] y^3 
    \nonumber \\
    &\quad + \Bigl[ -3 \AA^3 - 3 \AA^4 + 3 \AA^5 - 27 \AA \tilde{r} - 132 \AA^3 \tilde{r} 
    \nonumber \\
    &\quad - 165 \AA^4 \tilde{r} - 27 \AA^5 \tilde{r} + 
    339 \AA \tilde{r}^2 + 306 \AA^3 \tilde{r}^2 
    \nonumber \\
    &\quad + 339 \AA^4 \tilde{r}^2 + 27 \AA^5 \tilde{r}^2 + 
    27 (1 - 9 \tilde{r}) \tilde{r}^2 
    \nonumber \\
    &\quad + 108 \AA \tilde{r}^3 + 108 \AA^3 \tilde{r}^3 - 243 \AA^4 \tilde{r}^3 
    \nonumber \\
    &\quad - 3 \AA^2 \Bigl( -1 + \tilde{r} \bigl[ 55 + 6 \tilde{r} (-17 + 33 \tilde{r}) \bigr] \Bigr) \Bigr] y^2 
    \nonumber \\
    &\quad + \Bigl[ -6 \AA^2 - 12 \AA^3 - 6 \AA^4 - 6 \AA \tilde{r} + 18 \AA^3 \tilde{r} 
    \nonumber \\
    &\quad - 6 \AA^4 \tilde{r} + 54 \AA \tilde{r}^2 + 54 \AA^3 \tilde{r}^2 
    \nonumber \\
    &\quad - 18 \AA^2 \tilde{r} (-1 + 6 \tilde{r}) \Bigr] y + \Bigl[ - \AA - 3 \AA^2 - 3 \AA^3 
    \nonumber \\
    &\quad - \AA^4 + 9 \AA \tilde{r} - 9 \AA^2 \tilde{r} + 9 \AA^3 \tilde{r} \Bigr]
\end{align}
\endgroup
At $\tilde{r} = 0$, it holds
\begin{align}
    f(y, \tilde{r}) &= \AA (1 + \AA + 2 \AA y)^3 (-1 + \AA y^2)^3
    \overset{!}{=} 0
\end{align}
and we recover the critical BBP point
\begin{align}
    y_0 = \frac{1}{\sqrt{\AA}} \quad \iff \quad \vartheta^2_{\mathrm{BBP}} = \frac{\sigma^2}{\sqrt{\AA}} \ .
\end{align}
Now, we conduct a Taylor expansion in both variables around the point $(y_0, 0)$,
\begin{align} 
    f(y, \tilde{r}) 
    &= f(y_0, 0) + \frac{\partial f}{\partial y} (y_0, 0) \cdot (y - y_0) + \frac{\partial f}{\partial \tilde{r}} (y_0, 0) \cdot \tilde{r} 
    \nonumber \\
    &\quad + \frac{1}{2!} \frac{\partial^2 f}{\partial^2 y} (y_0, 0) \cdot (y - y_0)^2 + \frac{1}{2!} \frac{\partial^2 f}{\partial^2 \tilde{r}} (y_0, 0) \cdot \tilde{r}^2 \nonumber \\
    &\quad + \frac{\partial^2 f}{\partial y \partial \tilde{r}} (y_0, 0) \cdot (y - y_0) \tilde{r} 
    \nonumber \\
    &\quad + \frac{1}{3!} \frac{\partial^3 f}{\partial^3 y} (y_0, 0) \cdot (y - y_0)^3 + \dotsc
\end{align}
The constant term $f(y_0, 0)$ simply vanishes with the above considerations; similarly, we find
\begin{align}
    \frac{\partial f}{\partial y} (y_0, 0) = 0 \ , \qquad 
    \frac{\partial^2 f}{\partial^2 y} (y_0, 0)  = 0 \ .
\end{align}
Because we seek solutions $y = y(\tilde{r})$ to $f(y, \tilde{r}) = 0$, we expect a small change away from $\tilde{r} = 0$ to impact the solution as a small deviation to $y_0$ in the form of a power series w.r.t. $\tilde{r}$.
We let $y - y_0 = C_1 \tilde{r}^{k} + C_2 \tilde{r}^{2k} + \mathcal{O}(\tilde{r}^{3k})$ with some $k > 0$.
We get 
\begin{align} 
    f(y, \tilde{r}) 
    &= \frac{\partial f}{\partial \tilde{r}} (y_0, 0) \cdot \tilde{r} 
    \nonumber \\
    &\quad + \frac{1}{3!} \frac{\partial^3 f}{\partial^3 y} (y_0, 0) \cdot \Bigl( C_1 \tilde{r}^{k} + \mathcal{O}(\tilde{r}^{2k}) \Bigr)^3 \nonumber \\
    &\quad + \mathcal{O}(\tilde{r}^{1 + k}) + \mathcal{O}(\tilde{r}^{2}) \ .
\end{align}
The first non-vanishing terms, where ``first'' corresponds to the lowest orders $\tilde{r}^{3k}$ and $\tilde{r}$, are
\begin{subequations}
\begin{align}
    \frac{1}{3!} \frac{\partial^3 f}{\partial^3 y} (y_0, 0) 
    &= 8 (1 + \sqrt{\AA})^6 \AA^{5/2} \ ,
    \\
    \frac{\partial f}{\partial \tilde{r}} (y_0, 0) 
    &= - 27 (1 + \sqrt{\AA})^8 \ .
\end{align}
\end{subequations}
We collect the terms of the smallest overall order.
In the fashion of perturbation theory, the corresponding coefficient must vanish.
As the above derivatives are non-zero on their own, $\tilde{r}^{3k}$ and $\tilde{r}$ must be of the same order -- following the arguments of the method of dominant balance -- such that $3 k = 1$, or $k = \frac{1}{3}$.
Inserting $k$ with its specified value back into the series gives
\begin{align}
    f(y, \tilde{r}) 
    &= \biggl( \frac{\partial f}{\partial \tilde{r}} (y_0, 0) + \frac{1}{3!} \frac{\partial^3 f}{\partial^3 y} (y_0, 0) \cdot C_1^3 \Biggr) \tilde{r} 
    + \mathcal{O}(\tilde{r}^{4/3})
\end{align}
whereby $f(y,\tilde{r}) = 0$ now gives
\begin{align}
    C_1 
    &= \Biggl[ - \frac{\frac{\partial f}{\partial \tilde{r}} (y_0, 0)}{\frac{1}{3!} \frac{\partial^3 f}{\partial^3 y} (y_0, 0)} \Biggr]^{1/3}
    = \frac{3}{2} \frac{(1 + \sqrt{\AA})^{2/3}}{\AA^{5/6}}
\end{align}
In total, the low-rank scaling is
\begin{align}
    \frac{\vartheta^2_{\mathrm{crit}}}{\sigma^2} - \frac{\vartheta^2_{\mathrm{BBP}}}{\sigma^2}
    &= \frac{3}{2} \frac{(1 + \sqrt{\AA})^{2/3}}{\AA^{5/6}} \cdot \tilde{r}^{1/3} + \mathcal{O}(\tilde{r}^{2/3})
\end{align}
for the squared signal singular value.
If we were to transfer that scaling to the signal singular value itself, we would have
\begin{align}
    \Abs{\frac{\vartheta_{\mathrm{crit}}}{\sigma}}
    &= \sqrt{\frac{\vartheta_{\mathrm{BBP}}^2}{\sigma^2} + \biggl[ \frac{3}{2} \frac{(1 + \sqrt{\AA})^{2/3}}{\AA^{5/6}} \cdot \tilde{r}^{1/3} + \mathcal{O}(\tilde{r}^{2/3}) \biggr]}
    \nonumber \\
    &= \Abs{\frac{\vartheta_{\mathrm{BBP}}}{\sigma}} \biggl( 1 + \frac{3}{4} \frac{\sigma^2}{\vartheta^2_{\mathrm{BBP}}} \frac{(1 + \sqrt{\AA})^{2/3}}{\AA^{5/6}} \cdot \tilde{r}^{1/3} 
    \nonumber \\
    &\quad + \mathcal{O}(\tilde{r}^{2/3}) \biggl)
\end{align}
such that the scaling $\tilde{r}^{1/3}$ remains at the lowest order 
\begin{align}
    \Abs{\frac{\vartheta_{\mathrm{crit}}}{\sigma}} - \Abs{\frac{\vartheta_{\mathrm{BBP}}}{\sigma}}
    &= \frac{3}{4} \frac{(1 + \sqrt{\AA})^{2/3}}{\AA^{7/12}} \cdot \tilde{r}^{1/3} + \mathcal{O}(\tilde{r}^{2/3}) \ .
\end{align}

\paragraph{Algebraic Approach for Square Matrices.} 
For square matrices, the discriminant $\Delta$ of the Green function polynomial factors into a fixed power of $z$ and a cubic polynomial in $z$ (previously quintic).
Given the rescaled variable $y = \vartheta^2 / \sigma^2$, it holds
\begin{align}\label{eq:discriminant-A1}
    \Delta_{\mathcal{A}=1} 
    &= \sigma^{36} y \biggl( \frac{z}{\sigma^2} \biggr)^{\! 6} \biggl\{ 4 y^2 \biggl( \frac{z}{\sigma^2} \biggr)^{\! 3} + \Bigl[ -8 y + 36 \tilde{r} y  
    \nonumber \\
    &\quad - 27 \tilde{r}^2 y - 32 y^2 + 12 \tilde{r} y^2 - 8 y^3 \Bigr] \biggl( \frac{z}{\sigma^2} \biggr)^{\! 2} 
    \nonumber \\
    &\quad + \Bigl[ 4 - 4 \tilde{r} + 48 y - 196 \tilde{r} y + 88 y^2 - 124 \tilde{r} y^2 
    \nonumber \\
    &\quad + 48 y^3 - 60 \tilde{r} y^3 + 4 y^4 + 48 \tilde{r}^2 y (3 + y) \Bigr] \frac{z}{\sigma^2} 
    \nonumber \\
    &\quad + \Bigl[ -16 + 16 \tilde{r} - 64 y + 192 \tilde{r} y - 96 y^2 
    \nonumber \\
    &\quad + 352 \tilde{r} y^2 + 256 \tilde{r}^3 y^2 - 64 y^3 + 192 \tilde{r} y^3 - 16 y^4 
    \nonumber \\
    &\quad  + 16 \tilde{r} y^4 - 128 \tilde{r}^2 y (1 + 4 y + y^2) \Bigr] \biggr\} \ .
\end{align}
The lower boundary at the origin stems from the $z^6$ term.
For the third-order polynomial inside the square brackets, we expect one definite real root $z / \sigma^2$ for the upper boundary.
Large signals should give two more real roots for the inner boundaries.
As we seek the point where two of such roots coincide, we look at the discriminant of this polynomial of degree 3.
A general third-order polynomial $ax^3 + bx^2 + cx + d$ has the discriminant 
\begin{align}
    \tilde{\Delta} = 18 abcd - 4b^3 d + b^2 c^2 - 4 a c^3 - 27 a^2 d^2 \ .
\end{align}
Here, it evaluates to 
\begin{align}
    \tilde{\Delta} &= 16 \tilde{r} y^2 (1 - \tilde{r} + y)^2 \Bigl(108 \tilde{r}^2 y + 8 (-1 + y)^3 \nonumber \\
    &\quad + 9 \tilde{r} (1 - 14 y + y^2) \Bigr)^3 \ .
\end{align}
For non-zero signals, here $y > 0$, this expression is zero when the latter bracket vanishes. 
Again, we obtain a cubic equation, but now for $y$. 
For small $\tilde{r} < 4/5$, the real and physically relevant solution reads
\begin{align}
    y
    &= 1 + \frac{3}{8} \beta(\tilde{r})^{1/3} + \frac{2592 \tilde{r} - 
    2511 \tilde{r}^2}{216 \, \beta(\tilde{r})^{1/3}} - \frac{3 \tilde{r}}{8} \ ,
\end{align}
where 
\begin{align}
    \beta(\tilde{r}) &:= 128 \tilde{r} - 176 \tilde{r}^2 + 47 \tilde{r}^3 \nonumber  \\
    &\quad + 16 \tilde{r} \sqrt{64 - 304 \tilde{r} + 540 \tilde{r}^2 - 425 \tilde{r}^3 + 125 \tilde{r}^4} \ .
\end{align}
The critical BBP point lies at $\vartheta_{0}^2 = \sigma^2 / \sqrt{\mathcal{A}} = \sigma^2$, or $y_{0} = 1$, for $\mathcal{A} = 1$.
Thus, we get 
\begin{align}
    \frac{\vartheta^2}{\sigma^2} - \frac{\vartheta_{\mathrm{BBP}}^2}{\sigma^2}
    &= \frac{3}{8} \beta(\tilde{r})^{1/3} + \frac{2592 \tilde{r} - 
    2511 \tilde{r}^2}{216 \, \beta(\tilde{r})^{1/3}} - \frac{3 \tilde{r}}{8} \ .
\end{align}
We expand $\beta$ in terms of $\tilde{r}$ in this low-rank regime, which gives $\beta = 128\tilde{r} + 16 \cdot 8\tilde{r} + \mathcal{O}(\tilde{r}^2) = 256 \tilde{r} + \mathcal{O}(\tilde{r}^2)$ and therefore yields 
\begin{align}
    \frac{\vartheta^2}{\sigma^2} - \frac{\vartheta_{\mathrm{BBP}}^2}{\sigma^2} 
    &= \frac{3}{2^{1/3}} \, \tilde{r}^{1/3} + \mathcal{O}(\tilde{r}^{2/3}) \ .
\end{align}
This directly shows that the extensive BBP phase transition adheres to the observed scaling law of $\tilde{r}^{1/3}$, and it is consistent with the perturbative approach from above.

\section{Details on the Full-Rank Limit}
\label{app:full-rank-limit}
The full-rank limit can be dealt with similarly to the low-rank approximation. The self-consistency equation \eqref{eq:self-consistency-with-sum} may be split into a major contribution
\begin{align}\label{eq:full-rank-self-con}
    G(z) = \frac{1}{z \bigl[ 1 - \frac{\sigma^2}{\AA} \, G(z) \bigr] - \sigma^2 (1 - 1/\AA) - \frac{\vartheta^2}{1 - \frac{\sigma^2}{\AA} \, G(z)}} \ ,
\end{align}
which is now containing the signal, and a minor contribution from the noise-only part
\begin{align}\label{eq:full-rank-outlier-condition-1}
    0 = z \biggl[ 1 - \frac{\sigma^2}{\AA} \, G(z) \biggr] - \sigma^2 (1 - 1/\AA)
\end{align}
responsible for exceptions to the rule $-$ the outliers. They can be determined without explicit knowledge of the Green function. The Blue function, the functional inverse of $G$, can be used instead. Let us denote it by $B_{\mathrm{fr}}$ in this full-rank case. Eq.\,\eqref{eq:full-rank-outlier-condition-1} is then equivalent to
\begin{align}\label{eq:full-rank-outlier-condition-2}
    G(z) &= \frac{1}{\frac{\sigma^2}{\AA} - \vartheta^2} \qiff z = B_{\mathrm{fr}} \biggl( \frac{1}{\frac{\sigma^2}{\AA} - \vartheta^2} \biggr) \ .
\end{align}
Additionally, Eq.\,\eqref{eq:full-rank-self-con} yields the explicit form of $B_{\mathrm{fr}}$ via $z = B_{\mathrm{fr}}(G(z))$:
\begin{align}
    B_{\mathrm{fr}}(y) &= \frac{1}{1 - \frac{\sigma^2}{\AA} \, y} \biggl( \frac{1}{y} + \sigma^2 (1 - 1/\AA) + \frac{\vartheta^2}{1 - \frac{\sigma^2}{\AA} \, y} \biggr) \ .
\end{align}
The spectrum's boundaries are characterized by infinite slopes, hence any $y$ with $B'_{\mathrm{fr}}(y) = 0$ yields a boundary $B(y)$. Inserting $y = \inv{(\sigma^2 / \AA - \vartheta^2)}$ from Eq.\,\eqref{eq:full-rank-outlier-condition-2} into $B'_{\mathrm{fr}}(y) = 0$ reveals a value for the critical signal at which the outlier is exactly at a boundary. 
\begin{align}
    \vartheta^2_{\mathrm{crit}} &= \frac{\sigma^2}{\AA} \Bigl( 1 + \sqrt{\AA - 1} \Bigr) \ .
\end{align}
It is, in fact, the lower boundary, because larger signals shift the major bulk away, giving rise to separate delta peaks from the noise-only contribution. Similarly to the BBP phase transition, they are described by
\begin{align}
    x_0 &= \begin{cases}
        \sigma^2 (1 - 1/\AA) \bigl( 1 - \frac{\sigma^2}{\AA} \frac{1}{\vartheta^2} \bigr) \ , \ &\vartheta^2 > \frac{\sigma^2}{\AA} \bigl( 1 + \sqrt{\AA - 1} \bigr) \\
        \sigma^2 (1 - 1/\AA) \frac{\sqrt{\AA - 1}}{1 + \sqrt{\AA - 1}} \ , \ &\mathrm{else}.
    \end{cases}
\end{align}
only now with every of the $q - r \ll q$ noise-only eigenvalues at a common value.

\end{document}